Axel Oberschelp/Stephan Stahlschmidt

# Größe als Erfolgsgarant?

Zur Bedeutung der Organisationstruktur für die Einwerbung von Drittmitteln der Deutschen Forschungsgemeinschaft

**Research funding through third-party financing is of considerable importance for the German science system. The funds of the German Research Foundation (DFG) serve as the central focus due to the high reputation of the foundation. However, it has not been clarified yet to what extent the chances of successfully acquiring these funds depend on the structure of the university as an institution. The present study analyses DFG funding in the context of university research and examines the role of organisational conditions in the acquisition of funding. Several factors, such as *size* of the institution, *equipment*, and *teaching activities*, are analysed. The empirical study focuses on four subjects and investigates the correlation between funding success and conditional factors using a Bayesian approach. Results reveal the considerable relevance of the factors *size* as well as *provision of academic and non-academic personnel*. This implies that the organisational conditions are to be taken into account while evaluating third-party financing success.**

Seit etlichen Jahren steigt die Bedeutung von Forschungsevaluationen, wobei die Verwendung von Informationen zur Finanzierung der Forschung über Drittmittel eine wichtige Rolle spielt. Mit Blick auf die Praktiken einer Performanzmessung von Universitäten ist festzustellen, dass Drittmittelindikatoren bzw. Kennzahlen Bestandteile der überwiegenden Mehrzahl der eingesetzten Instrumente (z.B. Leistungsorientierte Mittelvergabe, Rankings) sind (Jansen/Görtz/Heidler 2015). Auschlaggebend für ein erfolgreiches Abschneiden von Universitäten bei einem Vergleich eingeworbener Drittmittel sind die Aktivitäten ihrer Mitglieder, insbesondere Umfang und Qualität der Antragstellungen sind entscheidend. Zunehmend stellt sich jedoch die Frage, welchen Beitrag Organisation hierbei leisten, sei es durch aktive Maßnahmen der Forschungsförderung, sei es durch ihre strukturelle Verfasstheit. Die hier vorgelegte Studie hat ein besseres Verständnis dieser Zusammenhänge zum Ziel und leistet damit einen Beitrag zu der in Deutschland intensiv geführten Diskussion um die Nutzung und Aussagekraft von Drittmittelindikatoren (vgl. Baier 2017; Winterhager 2015; Gerhards 2013).

## 1    Fragestellung und Forschungsstand

Der Wissenschaftsrat (WR) hob im Rahmen seines Forschungsratings, das von 2004 bis 2012 stattfand, hervor, dass eine Bewertung der Leistungsfähigkeit von universitären (Teil-)Organisationen insbesondere auf Kontextinformationen angewiesen sei, die eine Einordnung der gemessenen Ergebnisse ermöglichen. In vier Pilotstudien wurden neben Output-Daten zusätzlich Informationen zu Personalressourcen und Lehrtätigkeiten der bewerteten Einheiten erhoben, um „die Besonderheiten bestimmter Forschungsbereiche" in eine „kontextualisierte Bewertung" einfließen zu lassen (Wissenschaftsrat 2012). Im Mittelpunkt dieser Studie steht die Frage, welche Strukturelemente von Universitäten bedeutsam dafür sind, erfolgreich Drittmittel der Deutschen Forschungsgemeinschaft (DFG) zu akquirieren. Dabei ist eine Analyse von statistisch nachweisbaren Zusammenhängen zwischen dem Erfolg bei der Einwerbung von DFG-Drittmitteln (abhängige Variable) und organisationsbedingten Faktoren (unabhängige Variablen) Gegenstand des empirischen Teils. Als organisationsbedingte Faktoren werden Größe, Ausstattungsmerkmale (wissenschaftliches und nichtwissenschaftliches Personal), Lehraktivitäten und Kooperationsmöglichkeiten mit anderen Forschungseinrichtungen innerhalb und außerhalb der eigenen Organisation berücksichtigt. Die vorliegende Studie untersucht insbesondere Effekte, die mit Blick auf eine Übertrag- und Anwendbarkeit des theoretischen Konzepts New Public Management (NPM) auf den Hochschulbereich bedeutsam sind. Nach diesem Konzept konkurrieren Universitäten als zunehmend wettbewerbsfähige Akteure um Mittel (Enders/Kehm/Schimank 2015), wobei der Wettbewerb zusätzliche Leistungsreserven und -potenziale mobilisieren soll. Für die künftige hochschulpolitische Relevanz dieses Konzepts ist die Chancengleichheit bei diesen wettbewerblichen Verfahren von erheblicher Bedeutung. Auch nach einer nun bereits längeren Phase der Reformen und Restrukturierungen ist jedoch noch immer umstritten, ob NPM ein adäquater Reformansatz für den Hochschulbereich ist.

Vergleichsweise wenige Studien haben sich bislang mit der Frage befasst, wie die Leistungen von Universitäten durch organisatorische Rahmenbedingungen beeinflusst werden. Insbesondere der Zusammenhang zwischen der Größe der untersuchten (Teil-)Organisationen und deren Leistungsfähigkeit stand im Mittelpunkt empirischer Analysen. Sowohl die Studien von Lepori et al. (2015) als auch von Dohmen (2015) stellen eine positive Korrelation zwischen der Größe fachlicher Einrichtungen und der Einwerbung von Drittmitteln zur Forschungsfinanzierung fest. Im Gegensatz dazu haben die Studien von Auspurg et al. (2008) und Carayol und Matt (2004) keinen positiven Einfluss des Faktors Größe feststellen können, vielmehr konnte letztgenannte Studie hohe Forschungsleistungen insbesondere bei kleinen Einheiten identifizieren. Auch eine neuere Studie im deutschen Kontext berücksichtigt, bezogen auf den Publikationsoutput von Professuren in der Psychologie, den Einfluss organisatorischer Faktoren. Die Autor*innen stellen einen positiven Einfluss von Ausstattungsmerkmalen auf den Forschungsoutput fest, weisen jedoch darauf hin, dass vor allem individuelle und disziplinspezifische Faktoren die Forschungstätigkeit beeinflussen (Rathmann/Mayer 2017). Weitere Faktoren neben der Größe, deren Zusammenhang mit Forschungsleistungen bzw. dem Umfang eingeworbener Drittmittel untersucht wurden, sind die Reputation von Organisationen, Netzwerkstrukturen, Ausstattungsmerkmale sowie die Intensität von Lehraktivitäten. Auch hier ist die Befundlage jedoch keineswegs eindeutig (Grözinger/Fromm 2013; Hornbostel/Heise 2006; Hornbostel 1997).

## 2 Indikatoren[1] zur Drittmittelförderung und zum deutschen Wissenschaftssystem

Im Zuge von NPM wurde die Finanzierung von Forschungsaktivitäten verstärkt an wettbewerblich vergebene Mittel gekoppelt, u.a. um Anreize für Leistungssteigerungen zu setzen. Daten des Statistischen Bundesamtes weisen für Universitäten im Jahr 2016 eine Drittmittelquote von 46% aller Ausgaben für Forschung und Entwicklung aus, während diese Quote im Jahr 2000 noch bei 36% lag (Statistisches Bundesamt 2018; 2003). Auch in absoluten Zahlen ist die Zunahme der drittmittelbasierten Forschungsförderung bemerkenswert. Im deutschen Wissenschaftssystem wird insbesondere den von der DFG bewilligten Mitteln großes Gewicht beigemessen, die ein qualitativ anspruchsvolles Begutachtungsverfahren durchlaufen müssen (Gerhards 2013, S. 24; Boer/Enders/Schimank 2008, S. 48). Die Bedeutung von Drittmitteln für die Forschungsfinanzierung sind ein Grund für den als flächendeckend zu bezeichnenden Einsatz von Drittmittelindikatoren im deutschen Hochschul- und Wissenschaftssystem (Jansen/Görtz/Heidler 2015). Aber auch die leichte Verfügbarkeit von Daten zu Drittmitteln und deren vermeintlich gute, hochschulübergreifende Vergleichbarkeit sind Gründe für die hohe Prävalenz und Akzeptanz von Drittmittelindikatoren.

---

[1] Im Folgenden ist einheitlich von Indikatoren die Rede. Zur Unterscheidung von Kennzahlen und Indikatoren siehe Hornbostel (2001b).

In der Praxis der Leistungsmessung wird Erfolg bei der Einwerbung von Drittmitteln häufig mit Forschungserfolg gleichgesetzt (Lenz/Raßer 2012). Informationen zur Höhe eingeworbener Mittel werden als Stellvertreterdaten verwendet, um Forschungserfolg in metrischer Form mittels einer operationalisierbaren Größe abzubilden. Ein solches Vorgehen ist jedoch nur dann gerechtfertigt, wenn ein klar erkennbarer, direkter Zusammenhang zwischen der Höhe eingeworbener Drittmittel und der Quantität oder Qualität des Outputs von Forschungsaktivitäten, bspw. Publikationen, vorläge. Ein solcher unmittelbarer Zusammenhang ist jedoch nicht zweifelsfrei nachzuweisen. Jansen et al. (2007) haben in ihrer auf Forschungsgruppen bezogenen Studie vielmehr festgestellt, dass von einem abnehmenden Grenznutzen ausgegangen werden kann: Nur bis zu einer bestimmten Schwelle des Drittmittelvolumens nimmt die Publikationstätigkeit zu, bei Einwerbung von Drittmitteln über diese Schwelle hinaus ist hingegen eine Abnahme der Publikationstätigkeit festzustellen (vgl. Krempkow 2017). Auch Schmoch et al. (2010) kommen zu dem Ergebnis, dass der Zusammenhang von Drittmittel-Input und Publikations-Output nicht eindeutig zu belegen ist und Drittmittel als alleiniger Indikator zur Messung von Forschungsleistungen nicht geeignet sind.

Die Höhe der bewilligten Drittmittel indiziert in erster Linie den Erfolg der Antragsteller*innen bei der Einwerbung dieser Mittel. Im Hinblick auf den Forschungsprozess kann von einer Ausstattungsvariablen, einem Produktionsfaktor für die Ermöglichung von Forschung gesprochen werden (Gerhards 2013), wobei fachspezifisch sehr unterschiedliche Finanzierungsbedarfe in Anschlag zu bringen sind. Unter der Annahme, dass die gewährten Mittel tatsächlich für Forschungszwecke verwendet werden, sind Drittmittel darüber hinaus geeignet, den Umfang der Forschungsaktivitäten zu indizieren. Wenn – wie dies bei den Drittmitteln der DFG der Fall ist – ein qualitätssicherndes Begutachtungsverfahren mit der Mittelvergabe verbunden ist, wird daraus gebildeten Drittmittelindikatoren zugutegehalten, sie würden indirekte Qualitätsurteile enthalten (Hornbostel 2008). Derartige Zuschreibungen der Peers beziehen sich allerdings zunächst nur auf die Qualität der Antragstellung. Eine im Jahr 2015 veröffentlichte Studie zur Bewilligungspraxis des Austrian Science Fund (FWF) hat zwar gezeigt, dass ex-ante-Bewertungen von Forschungsvorhaben mit ex-post-Bewertungen der tatsächlich durchgeführten Projekte korrelieren (Mutz/Bornmann/Daniel 2015). Dies spricht dafür, Bewilligungsentscheidungen als Prädiktor für zu erwartende Forschungserfolge heranziehen. Andererseits haben ältere Studien nachgewiesen, dass bei Gutachterentscheidungen, etwa der DFG, durchaus auch sachfremde Aspekte wie z.B. das Senioritätsprinzip eine Rolle spielen (Olbrecht 2013; Klein/Kraatz/Hornbostel 2012).

# 3  Datenbasis und univariate Analysen

Da auf die Gesamtorganisation Universität bezogenen empirischen Analysen nur geringe Aussagekraft zukommt, sind Fächer Gegenstand dieser Untersuchung. Bei ihrer Auswahl war ein zentraler Aspekt, dass die Finanzierung von Forschungsvorhaben durch Fördermittel der DFG von beträchtlicher Bedeutung ist (vgl. Hornbostel 2001a, S. 532). Für die Forschungsfinanzierung der hier untersuchten Fächer

- Geschichte, Archäologie (im Folgenden kurz *Geschichte*),

- Psychologie (*Psychologie*),
- Physik, Astronomie (*Physik*) und
- Maschinenbau, Verfahrenstechnik (*Maschinenbau*)

ist dies zutreffend. Für diese Fächer lassen sich zudem beträchtliche, fachbezogene Unterschiede in der Drittmittelausstattung feststellen, wodurch eine relevante Varianz der Untersuchungsobjekte gegeben ist. Schließlich handelt es sich innerhalb der zuzuordnenden Fächergruppen um vergleichsweise große Fächer, was u.a. den Nachweis von Skaleneffekten erleichtert.

Die Datenbasis dieser Studie sind standortbezogene Angaben der DFG zum Fördervolumen (gerundete Summenwerte für den Zeitraum 2014-2016), entnommen dem „Förderatlas" der DFG (Deutsche Forschungsgemeinschaft 2018) sowie Daten des Statistischen Bundesamtes zu Beschäftigten (Auswertungsjahr 2016) und Studierenden (Wintersemester 2015/16). Für die Berechnung einzelner Kontextvariablen werden zudem online verfügbare Daten der DFG-Datenbank GEPRIS sowie georeferenzielle Daten des Bundesinstituts für Bau-, Stadt- und Raumforschung (BBSR) herangezogen. Die Zusammenführung von Daten der DFG und des Statistischen Bundesamtes, denen unterschiedliche Fachsystematiken zugrunde liegen, hat u.a. zur Folge, dass die hier untersuchten Fächer Artefakte darstellen, welche jedoch die an den Universitäten tatsächlich vorzufindenden Teilorganisationen näherungsweise repräsentieren. Die folgenden Variablen (Leistungs- und Kontextvariablen) gehen in das Modell ein:

**Tabelle 1: Übersicht der Modellvariablen**

| Leistungsvariable | |
|---|---|
| dfg-drm_prof | DFG-Drittmittel, normiert auf die Anzahl der Professor*innen |
| **Kontextvariablen** | |
| wp | Größe des Faches (Anzahl Professor*innen und sonstiges wissenschaftliches Personal) |
| swp_prof | Ausstattung mit wissenschaftlichem Personal (sonstiges wissenschaftliches Personal je Professor*in) |
| nwp_prof | Ausstattung mit nichtwissenschaftlichem Personal (nichtwissenschaftliches Personal je Professor*in) |
| stud_wp | Umfang der Lehraktivitäten (Studierende je wissenschaftliches Personal) |
| stud_la | Anteil Studierende im Lehramt |
| stud_ms | Anteil Master-Studierende |
| koop_reg | regionales Kooperationspotential (fach- und standortspezifisches Potential für regionale Forschungskooperationen = Bestimmung des Näheverhältnisses zu anderen, fachlich vergleichbaren Einrichtungen unter Verwendung georeferentieller Daten, die Entfernungen werden in PKW-Fahrzeiten übersetzt und gehen in die Modellierung der Variable ein) |
| koop_int | internes Kooperationspotential (Potential für interdisziplinäre Kooperationen innerhalb der eigenen Organisation= Ermittlung fachspezifischer Kooperationsprofile orientiert an Forschungsvorhaben im Rahmen der DFG-Verbundforschung und Abgleich mit dem fachlichen Profil der untersuchten Universitäten) |

Die Frage nach Konzentrationseffekten bei der Vergabe von Fördermitteln auf einzelne Standorte ist für das Wissenschaftssystem insgesamt relevant. Eine Analyse der eingeworbenen Mittel ergibt, dass im Fach *Geschichte* die Konzentration der Mitteleinwerbung am größten ist: Auf 20% der Einrichtungen entfallen hier 71% der für das gesamte Fach bewilligten Mittel, während bei den anderen drei Fächern nur ca. 50% der Mittel auf diese obersten 20% entfallen. Für alle vier Fächer liegt die Konzentration der eingeworbenen Mittel dabei deutlich über der – hier nicht separat ausgewiesenen – Konzentration des wissenschaftlichen Personals, für die bezogen auf das 1. Quintil Werte zwischen 33% und 44% festzustellen sind.

Tabelle 2: Anteil der 20% einwerbungsstärksten Einrichtungen an den gesamten Einwerbungen des Faches

|  | 1. Quintil | | |
| --- | --- | --- | --- |
|  | 2011-2013 | 2014-2016 | Differenz |
| *Geschichte* | 67% | 71% | +4% |
| *Psychologie* | 51% | 55% | +4% |
| *Maschinenbau* | 53% | 53% | ±0% |
| *Physik* | 50% | 48% | -2% |

(DFG-Drittmittel, ohne Normierung auf die Anzahl der Professor*innen)

Der Vergleich mit den Ergebnissen der vorangegangenen Förderphase (2011-2013) ergibt für die Fächer *Geschichte* und *Psychologie* eine zunehmende Konzentration der Mittel. Dieser Befund weicht von den Berechnungen der DFG ab, die ausgehend von einer anderen Berechnungsmethode[2] abnehmende Konzentrationseffekte konstatiert (Deutsche Forschungsgemeinschaft 2018). Die univariate Analyse (Abbildung 1) der Leistungsvariable zeigt beträchtliche disziplinspezifische Niveauunterschiede sowie Unterschiede in der Verteilung auf die hier untersuchten Standorte. Im Fach *Geschichte* erhält ein beträchtlicher Anteil von mehr als 15% der Einrichtungen überhaupt keine DFG-Förderung, während alle Einrichtungen in der *Physik* eine DFG-Förderung erhalten.

---

[2] Die DFG vergleicht Rankingpositionen einzelner Einrichtungen, z.B. die Ränge 1 und 40, und berechnet für die einzelnen Zeitschnitte, um welchen Faktor das Bewilligungsvolumen der Spitzenposition höher ist als das der niedrigen Rangposition. Diese Methode, die nur Aussagen über die beiden jeweils betrachteten Einrichtungen zulässt, ist nach unserer Auffassung nicht geeignet, Verteilungen der Gesamtpopulation oder einzelner Teilgruppen zu analysieren.

**Abbildung 1: Verteilung (graue Fläche) der DFG-Fördermittel pro Professor mit begleitendem Median (graue durchgezogene Linie) und 25%/75%-Quartil (graue gestrichelte Linien). Für eine übersichtliche Darstellung wurde auf der Y-Achse eine zufällige Schwankung eingefügt.**

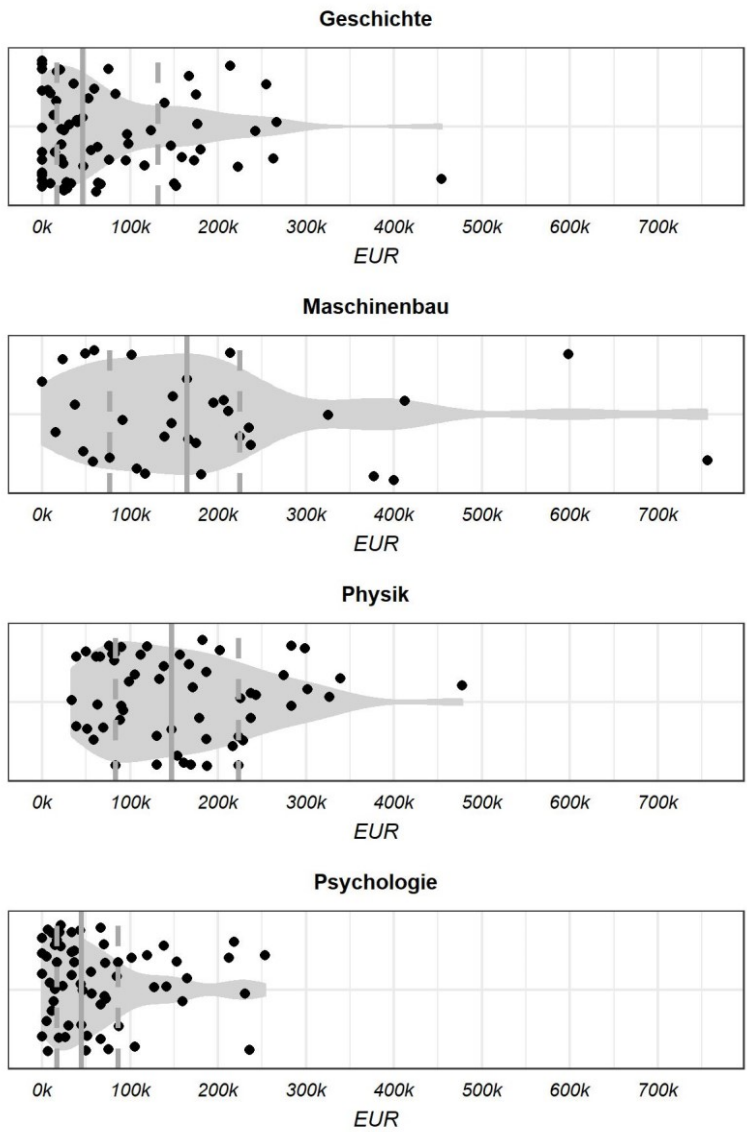

Die folgende Auswertung enthält zentrale statistische Maßzahlen zu den Ausprägungen der Leistungsvariable und der Kontextvariablen für die vier untersuchten Fächer.

**Tabelle 3: Übersicht über Verteilungen der Leistungs- und Kontextvariablen**

|  | Anzahl | Median | Min | Max | Anzahl | Median | Min | Max |
|---|---|---|---|---|---|---|---|---|
|  | Geschichte | | | | Psychologie | | | |
| *dfg-drm_prof* | 67 | 45.833 | 0 | 454.167 | 61 | 44.444 | 0 | 253.333 |
| *wp* | 67 | 21,8 | 2,1 | 100,0 | 61 | 34,7 | 4,0 | 87,3 |
| *swp_prof* | 67 | 1,6 | 0,5 | 2,9 | 61 | 2,5 | 0,2 | 4,7 |
| *nwp_prof* | 67 | 0,5 | 0,0 | 1,5 | 61 | 0,8 | 0,0 | 3,8 |
| *stud_wp* | 61 | 21,0 | 2,6 | 52,5 | 51 | 18,2 | 4,9 | 41,0 |
| *stud_la* | 61 | 32,1% | 0,0% | 100,0% | 56 | 0,0% | 0,0% | 100,0% |
| *stud_ms* | 59 | 16,8% | 3,2% | 100,0% | 49 | 30,3% | 0,0% | 60,3% |
| *koop_reg* | 67 | 10 | 1 | 23 | 61 | 9 | 1 | 29 |
| *koop_int* | 67 | 32 | 6 | 40 | 61 | 15 | 4 | 19 |
|  | Physik | | | | Maschinenbau | | | |
| *dfg-drm_prof* | 57 | 147.126 | 33.333 | 477.778 | 33 | 164.444 | 0 | 756.140 |
| *wp* | 57 | 54,8 | 9,0 | 154,6 | 33 | 93,4 | 1,0 | 426,7 |
| *swp_prof* | 57 | 2,2 | 0,5 | 4,4 | 32 | 3,7 | 0,1 | 8,4 |
| *nwp_prof* | 57 | 2,2 | 0,0 | 4,5 | 33 | 3,2 | 0,0 | 13,2 |
| *stud_wp* | 56 | 9,6 | 2,3 | 60,0 | 32 | 17,1 | 4,7 | 30,7 |
| *stud_la* | 55 | 10,0% | 0,0% | 40,3% | 33 | 0,0% | 0,0% | 5,8% |
| *stud_ms* | 55 | 23,3% | 4,3% | 49,6% | 33 | 33,0% | 0,0% | 100,0% |
| *koop_reg* | 57 | 16 | 0 | 34 | 33 | 20 | 2 | 48 |
| *koop_int* | 57 | 13 | 9 | 23 | 33 | 12 | 6 | 19 |

Bei der Größenvariable (*wp*) weist *Geschichte* den niedrigsten Median der hier betrachteten Fächer auf, die größten fachlichen Einrichtungen finden sich bei *Maschinenbau*. Im Median der Ausstattungsvariablen kommen die fachbezogenen Unterschiede wissenschaftlicher Produktion deutlich zum Ausdruck. Insbesondere bei der Variable *swp_prof* weisen *Geschichte* und *Psychologie* ein im Vergleich zu den naturwissenschaftlich-technischen Fächern deutlich niedrigeres Ausstattungsniveau auf. Aber auch bei der Ausstattung mit sonstigem wissenschaftlichen Personal (*swp_prof*) ist das Niveau in dem am besten ausgestatteten Fach *Maschinenbau* um den Faktor 2,4 höher als in dem Fach mit der niedrigsten Ausstattung (*Geschichte*). Hinsichtlich der Lehrtätigkeit (*stud_wp*) sind die Mediane aller Fächer mit Ausnahme von *Physik* in etwa vergleichbar. Die größten regionalen Kooperationspotentiale (*koop_reg*) liegen in den Fächern *Physik* und *Maschinenbau* vor. Grund hierfür sind zahlreiche fachliche Anknüpfungspunkte zu außeruniversitären Forschungseinrichtungen und Hochschulen für angewandte Wissenschaften. Große interne Kooperationspotentiale sind für *Geschichte* festzustellen. Dies ist ursächlich auf den Teilbereich Archäologie zurückzuführen, für den eine Vielzahl von Fächern, auch aus dem Bereich Naturwissenschaften und Technik, als Kooperationspartner in Frage kommen.

# 4 Empirischer Zusammenhang zwischen Drittmittelerfolg und organisationalen Spezifika der Fächer

Die zur Vorbereitung der Regressionsanalyse durchgeführten bivariaten Analysen zeigen insbesondere bei der Variable Größe (*wp*) deutliche (Spearman-Korrelationskoeffizient ≥ 0.5), teilweise kurvilineare Zusammenhänge zur abhängigen Variable, und zwar für alle Fächer außer *Psychologie*. Aber auch für dieses Fach ergibt sich ein Wert nahe 0.5. Die Variable *swp_prof* korreliert hingegen nur für das Fach *Maschinenbau* mit der abhängigen Variable. Die Ausstattung mit sonstigem wissenschaftlichem Personal ist hingegen ein Faktor von eher geringer Bedeutung für die übrigen Fächer. Bei der Ausstattung mit nichtwissenschaftlichem Personal (*nwp_prof*) sind für die Fächer *Geschichte* und *Psychologie* deutliche Korrelationen zur abhängigen Variable erkennbar. Während von regionalen Kooperationspotentialen (*koop_reg*) kein erkennbarer Einfluss ausgeht, besteht eine starke Korrelation zwischen der Ausprägung des internen Kooperationspotentials (*koop_int*) und der abhängigen Variable im *Fach Geschichte*. Keine Zusammenhänge sind hingegen für die Variable Lehrtätigkeit (*stud_wp*) zu erkennen. Zudem lassen sich Zusammenhänge der Kontextfaktoren untereinander festzustellen, welche den gewählten multivarianten Ansatz in Form eines Regressionsmodells begründen.

Um in einem solchen Ansatz potentielle disziplinäre Unterschiede aufzeigen zu können, wird ein lineares, gemischtes Bayesianisches Modell (Gelman et al. 2004; Bürkner 2017) basierend auf einem Hamiltonian Monte Carlo sampler (Homan/Gelman 2014; Carpenter et al. 2017) verwendet. Ein solcher Bayesianischer Ansatz zeichnet sich durch eine hohe Flexibilität in der Modellierung von (unsicheren) Zusammenhängen aus, während gleichzeitig zuvor bekannte Informationen explizit mittels der a-priori Wahrscheinlichkeit eingefügt werden können. Die resultierende a-posteriori Wahrscheinlichkeit ermöglicht eine Interpretation der Ergebnisse in Form von Wahrscheinlichkeitsaussagen, deren Interpretation im Vergleich zu frequentistischen p-Werten auf weniger strikten Annahmen beruht (s.a. Diskussion in Wasserstein/Schirm/Lazar 2019). Umgesetzt wird dieser Ansatz mittels eines log-normal hurdle Modells, um die rechtsschiefe Verteilung der Drittmittel und Beobachtungen ohne Drittmittel explizit einbeziehen zu können. Fehlende Werte zu Studierendenanteilen werden zuvor mittels multipler Imputation ersetzt. Der Erwartungswert der a-priori Wahrscheinlichkeit wird außer für die Lehrtätigkeit und den Anteil von Lehramtsstudierenden als positiv gesetzt, wobei die angewandte Normalverteilung aller a-priori Wahrscheinlichkeiten der erklärenden Variablen negative und positive Werte zulassen und somit flexibel definiert sind.

Basierend auf der Konvergenz der Markow-Ketten gibt Abbildung 2 die disziplinübergreifenden festen Effekte als a-posteriori Wahrscheinlichkeiten wieder. Beispielsweise wird die a-priori Verteilung des Einflusses der Größenvariable *wp* auf einen Mittelwert von 10% in einem durch einen Normalverteilung approximativ definierten Intervall von [-20%,50%] festgelegt. Nach Einbeziehung der erhobenen Daten in die Schätzung der a-posteriori Verteilung verschiebt sich diese mittlere Effektgröße auf den Mittelwert 1% mit einer sehr geringen Varianz. Da die Drittmittel in dieser Darstellung logarithmiert sind, können die Koeffizienten der erklärenden Variablen in Abbildung 2, soweit sie nicht allzu stark vom Wert „0" abweichen, nach Multiplikation mit dem Faktor „100" als approximative prozentuale Änderungen in den auf Professor*innen normierten Drittmitteleinwerbungen interpretiert werden.

**Abbildung 2: a-posteriori Wahrscheinlichkeiten der Koeffizienten für erklärende Variablen. Werte können, nach Multiplikation mit 100, approximativ als prozentuale Veränderungen interpretiert werden. Vertikale Nulllinie ist als visueller Ankerpunkt eingefügt.**

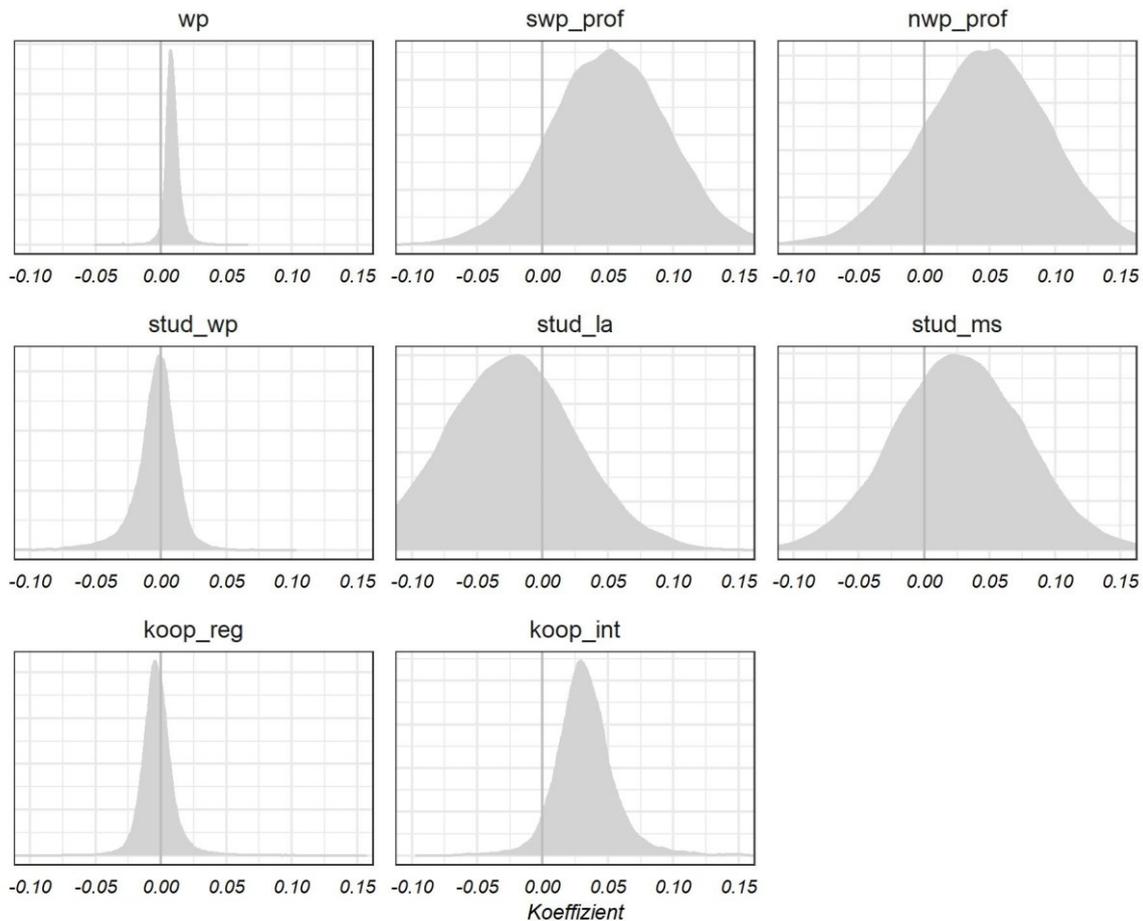

Die festen, disziplinübergreifenden Effekte unterscheiden sich hinsichtlich ihrer Einflussgröße, ihrer im Modell ermittelten Genauigkeit als auch ihrer fachspezifischen Ausprägung. Die Größenvariable (*wp*) weist den eindeutigsten Effekt auf, da die a-posteriori Wahrscheinlichkeitsverteilung kaum um den mittleren Wert 0,01 variiert. Entsprechend steigen die DFG Drittmittel pro Professor*in im Mittel für jede/n zusätzliche/n Wissenschaftler*in um 1%, d.h. umso mehr Mitglieder eine Untersuchungseinheit hat, desto mehr Drittmittel werben diese unter ansonsten gleichen Bedingungen ein. Die Wahrscheinlichkeit für einen positiven Einfluss beträgt dabei über 95%, während gleichzeitig die Wahrscheinlichkeit für eine Verdoppelung des Effekts auf 2% mit ca. 12% relativ gering ist. Disziplinäre Unterschiede sind für diesen Effekt nicht zu beobachten.

Die fixen Effekte der Ausstattungsvariablen (*swp_prof, nwp_prof*) weisen dagegen eine größere Bandbreite in der Wahrscheinlichkeitsverteilung auf. Der mittlere Koeffizientenwert liegt bei jeweils 0,05, also einer 5% höheren Effizienz in der Drittmitteleinwerbung für jede Erhöhung der Ausstattungsvariable um den Wert 1. Die Wahrscheinlichkeit für eine Effektverdoppelung auf eine 10% höhere Effizienz in der Drittmitteleinwerbung liegt bei unter 15% und die Wahrscheinlichkeit für einen negativen Effekt bei unter 20%. Insbesondere im Maschinenbau führen zusätzliche wissenschaftliche Mitarbeiter*innen zu steigenden Drittmitteleinwerbungen, während dieser Effekt zu einem geringeren Ausmaß in der Physik und Psychologie zu beobachten ist. Professor*innen im Fach Geschichte hingegen profitieren nicht von zusätzlichem wissenschaftlichem Personal, jedoch stärker als die drei anderen Disziplinen von zusätzlichem nicht-wissenschaftlichen Personal. Hierin kommt möglicherweise die Unterausstattung der Einrichtungen dieses Faches hinsichtlich dieser Personalkategorie (vgl. Tabelle 3) zum Ausdruck.

Dass sich ein zunehmender Umfang von Lehraktivitäten (*stud_wp*) zum Nachteil der Drittmitteleinwerbung auswirkt, kann nur für die Psychologie beobachtet werden, während im Maschinenbau und in der Physik positive Effekte zu beobachten sind. In der fächerübergrefenen Darstellung in Abbildung 2 führt dies zu einem neutralen, festen Effekt. Der Anteil Lehramts- (*stud_la*), bzw. Masterstudierender (*stud_ms*) führt ceteris paribus zu negativen, festen Effekten für Lehramtsstudierende, bzw. zu positiven, festen Effekten für Masterstudierende. Nur für die Psychologie kehren sich diese beiden Effekt um. Die Streuung der zugehörigen, festen Effekte weist allerdings auf eine beträchtliche Unsicherheit des Modells hinsichtlich des tatsächlichen Einflusses dieser Faktoren hin.

Die Kooperationsvariablen zeigen dagegen eine geringe Dispersion. Während für *koop_reg* ein sehr schwacher, negativer Effekt zu ermitteln ist, beträgt der mittlere Effekt für das intra-organisationale Kooperationspotential (*koop_int*) 0,03, d.h. je zusätzlicher Einheit des Kooperationspotentials ist mit einer Zunahme der Drittmitteleinwerbungen von 3% zu rechnen. Dieser positive Effekt lässt sich für alle vier Disziplinen beobachten. Die Wahrscheinlichkeit für einen negativen Effekt des intra-organisationalen Kooperationspotentials beläuft sich auf unter 7%.

# 5 Fazit

Die vorliegende Studie ist der Frage nachgegangen, inwieweit sich für Facheinheiten von Universitäten ein statistisch nachweisbarer Zusammenhang zwischen organisationalen Faktoren und dem Erfolg bei der Einwerbung von DFG-Drittmitteln feststellen lässt. Neben Größe und Ausstattungsmerkmalen wurden verschiedene Aspekte der Lehrtätigkeit sowie Kooperationspotenziale in die Analyse einbezogen. Auf einer deskriptiven Ebene haben die Ergebnisse die disziplinspezifischen Unterschiede in der Forschungsfinanzierung durch DFG-Drittmittel deutlich gemacht. Deutlich erkennbare Skaleneffekte bewirken die Variablen Größe (*wp*) sowie die Variablen zur personellen Ausstattung der Einrichtungen (*swp_prof, nwp_prof*). Die hierbei festzustellenden disziplinspezifischen Unterschiede sind teilweise beträchtlich. Die im vorangegangenen Abschnitt diskutierten Befunde zur Bedeutung von Lehraktivitäten für die Einwerbung von DFG-Drittmitteln ergänzen die zahlreichen, zum Teil disparaten Ergebnisse anderer Studien zum Verhältnis von Forschungs- und Lehraktivitäten (bspw. Menger 2016; Hattie/Marsh 2004). Während sich der Umfang der Lehraktivitäten je nach Fach unterschiedlich auswirkt, zeigen sich für die Anteile der Lehramts- bzw. Masterstudierenden Effekte, welche der erwarteten Richtung entsprechen: Bei einer Zunahme der eher praxisorientierten und wenig forschungsaffinen Lehramtsausbildung ist mit sinkenden Drittmitteleinwerbungen zu rechnen. Hingegen wirkt sich die Erhöhung des Anteils der forschungsorientierten Master-Ausbildung positiv auf die Höhe der Einwerbungen aus. Die nachgewiesenen Effekte sind allerdings schwach. Während ein von regionalen Kooperationspotenzialen ausgehender Einfluss auf die Höhe der eingeworbenen Drittmittel nicht nachweisbar ist, geht von organisationsinternen Kooperationspotenzialen ein verstärkender Effekt auf die Möglichkeiten der Drittmitteleinwerbung aus.

Aus den Ergebnissen dieser Studie ergeben sich wichtige Implikationen für die praktische Anwendung von leistungsmessenden Verfahren: So sollten bei einer Beurteilung der Höhe eingeworbener

Drittmittel die an den jeweiligen Standorten vorzufindenden Rahmenbedingungen berücksichtigt werden, dies gilt insbesondere für die Größe, für Ausstattungsmerkmale und für Möglichkeiten der organisationsinternen, interdisziplinären Kooperation. Dadurch könnten die Fairness der eingesetzten Verfahren sowie deren Akzeptanz auf Seiten der Verantwortlichen an den Universitäten erhöht werden.

Ausgelöst durch Reformen des NPM hat sich die Finanzierung von Forschungsaktivitäten in Deutschland, aber auch in anderen Ländern, in den letzten 30 Jahren stark verändert. So musste in den vergangenen Jahren ein immer größerer Teil der Mittel von den Forschenden kontinuierlich auf Projektebene eingeworben werden. Die vorliegende Studie hat gezeigt dass die Größe der fachlichen Einheiten hierbei ein relevanter Faktor ist.

Weitere, möglicherweise relevante Aspekte sind nicht Teil dieser Untersuchung, könnten allerdings Gegenstand künftiger Forschungen sein. So wurden die strukturellen Unterschiede in der Forschungsförderung, die sich aus der Inanspruchnahme unterschiedlicher Förderlinien der DFG ergeben, nicht berücksichtigt. Eine stärker differenzierende Analyse könnte neben der absoluten Höhe der Fordersumme auch die disziplinspezifisch variierenden Anteile von Einzelförderung, Verbundprojekten und Nachwuchsförderung in Rechnung stellen und mit den vorliegenden Erkenntnissen kombinieren. Die Organisationsstrukturen in den Blick nehmend wäre die Einbeziehung des organisatorischen Umfelds der fachlichen Einrichtungen innerhalb der Universitäten von Interesse, bspw. ob es sich um große, fachlich heterogene oder eher um kleine und fachlich homogene Fakultäten bzw. Fachbereiche handelt. Von Interesse wäre zudem, welchen Einfluss die Förderaktivitäten der Universitäten bei der Einwerbung von Mitteln haben. Die Möglichkeiten zu Forschungskooperationen konnten im Rahmen dieser Studie zwar in einer quantifizierenden Perspektive untersucht werden, bei diesem Zugang mussten aber Fragen nach der tatsächlichen Inanspruchnahme von Kooperationsmöglichkeiten und nach den Faktoren, die ihrer Nutzung im Wege stehen, offen bleiben. Eine Untersuchung dieser Aspekte steht mit Blick auf die Möglichkeiten ihrer Operationalisierung durch Bildung von Indikatoren allerdings vor beträchtlichen Herausforderungen. Eine vergleichbare Untersuchung zur Bedeutung des Faktors Größe für einen vor dem Einsetzen der NPM-Reformen gelegenen Zeitpunkt, wäre eine weitere vielversprechende Erweiterung der Perspektive dieser Studie und nähme die von diesen Reformen ausgehenden Effekte in den Blick.

# 6    Literatur